\documentstyle[preprint,aps,psfig,floats]{revtex}
\tighten
\begin{document}
\date{\today}
\draft
\title{
Is There A Phase Transition in the Isotropic Heisenberg\\
Antiferromagnet on the Triangular Lattice?}
\vspace{1in}
\author{W. Stephan\thanks{Present address: Bishop's University, Lennoxville, Qu\'{e}bec, 
Canada J1M 1Z7} and B.W. Southern}
\address{
Department of Physics and Astronomy\\
 University of Manitoba \\
Winnipeg, Manitoba, Canada R3T 2N2}
\maketitle
\begin{abstract}
   The phase diagram of the classical anisotropic (XXZ) Heisenberg model
on the 2-dimensional triangular lattice is investigated using Monte Carlo
methods. In the easy-axis limit,
two finite temperature vortex unbinding transitions have been observed. In the easy-plane limit, 
there also appear to be two distinct finite
temperature phase transitions which are very close in temperature. The upper transition 
corresponds to an Ising-like chirality ordering and the lower temperature transition corresponds 
to a Kosterlitz-Thouless vortex unbinding transition.  These
phase transition lines all meet at the Heisenberg point and provide strong evidence that
the isotropic model undergoes a novel finite temperature phase transition.
\end{abstract}
\vspace{0.5in}

\section {INTRODUCTION}

    The classical Heisenberg model on the two
dimensional triangular lattice with antiferromagnetic
nearest neighbour coupling is a geometrically frustrated system.
Kawamura 
and Miyashita\cite{1} have previously studied the possibility of a
Kosterlitz-Thouless-like\cite{2} transition in this system using Monte Carlo methods
and obtained evidence for a finite temperature transition driven by
the dissociation of vortices.  However, a transition at finite temperature
is considered by many to be unlikely for the Heisenberg model in two dimensions.
Dombre and Read\cite{3} mapped the continuum version of the model onto
a non-linear sigma model  (NL$\sigma$) and Azaria et. al.\cite{4}  used renormalization
group (RG) techniques to study the correlation length and effective
long-wavelength spin stiffness of this NL$\sigma$ at low temperatures.
Their results were consistent with a phase transition at $T_c=0$ with
an enlarged SO(3) symmetry. Indeed, Monte Carlo
simulations by Southern and Young\cite{5} of the spin stiffness seemed to provide partial 
support for this picture.
More recent work by Southern and Xu\cite{6} calculated the rigidity of the isotropic system against the
formation of free vortices at low temperatures. The vorticity stiffness
was found to be finite at low $T$ and disappeared abruptly near $T=0.31J$. This behaviour 
is consistent with a possible Kosterlitz-Thouless defect unbinding transition. It appears that the
stiffness and vorticity behave quite differently in the isotropic model.
The stiffness is zero on long length scales indicating that the two-spin
correlations decay exponentially at all finite $T$. However, the vorticity
indicates that an unbinding of topological defects occurs at a finite $T_c$.

\section{The Model and Results}

In an attempt to understand this difference in behaviour of the stiffness and vorticity, 
we have recently studied\cite{7} the following anisotropic model 
\begin{eqnarray}
 H = +J \sum_{i<j} [{ S_i^xS_j^x +  S_i^yS_j^y+ A S_i^zS_j^z}] 
\end{eqnarray}
where ${ S^\alpha_i , \alpha =x,y,z}$ represents a classical 3-component spin of 
unit magnitude
located at each site $i$ of a triangular lattice and the exchange interactions  are
restricted to nearest neighbour pairs of sites. The range $A>1$ corresponds to an
easy axis system. The ground state in this range exhibits two continuous symmetries which 
can be described in terms of two phase angles.\cite{1,7}
We found that the system undergoes two finite
temperature phase transitions associated with the unbinding of vortices related to these phase angles.

Information about the rigidity of the anisotropic system against fluctuations
 can be obtained from the spin wave stiffness and vorticity modulus.
The spin stiffness (helicity) tensor is given by the
second derivative of the free energy\cite{5,6,7} with respect
 to the twist angle about
a particular direction in spin space.   
Similarly, the vorticity\cite{6,7}
can be defined as the response of the spin system to an imposed twist 
about a given axis in spin space
along a closed path which encloses a vortex core. This is 
essentially the response of the system to an isolated vortex and can
be calculated as the second derivative of the free energy with respect
to the strength of the vortex, or winding number $m$, evaluated at $m=0$.
These two response functions can be calculated using standard Monte Carlo
methods for lattices of increasing size. We find that the spin stiffnesses and the corresponding 
vorticity moduli behave identically for the easy axis case $A>1$. In contrast, at the isotropic 
Heisenberg limit ($A=1$), the spin stiffness vanishes at large length scales but the vorticity 
moduli are non-zero at low T and vanish abruptly at a finite temperature. Similar work on the 
$xy$ model \cite{8} also indicated that the vorticity
and stiffness behave identically and that there are two phase transitions. These transitions 
are very close in temperature with the upper transition corresponding to an Ising-like transition 
and the lower to a Kosterlitz-Thouless transition. In the easy plane range of the present model, 
$-0.5<A<1$, the ground state
corresponds to a $120^0$ planar arrangement of the spins with the chirality\cite{9}
 on each triangle aligned along the $z$-axis. Recent Monte Carlo work by Capriotti et. al. \cite{10} 
in the range $0< A <1$  has also suggested two closely spaced transitions which are qualitatively 
the same as those in the $xy$ model.

\begin{figure}[t]
\centerline{\psfig{file=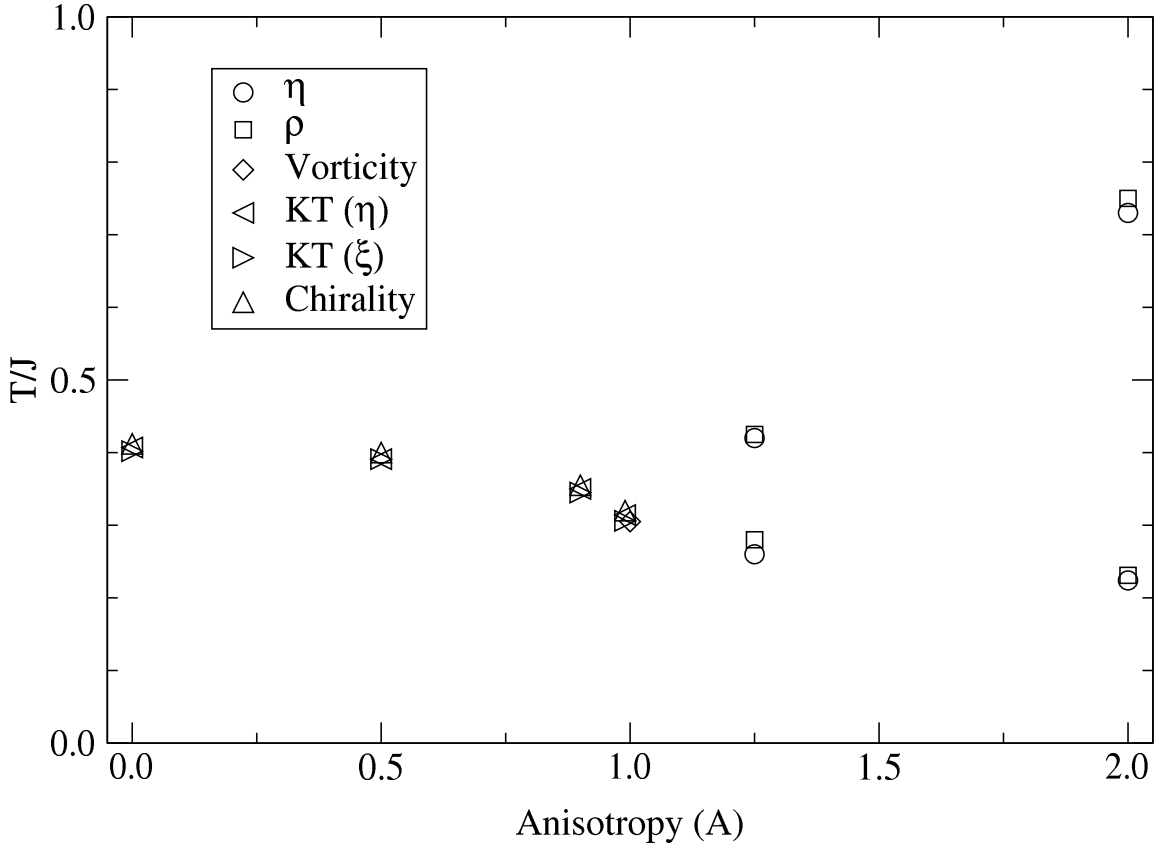,width=6.0truein,angle=-90}}
\caption{The transition temperatures $T_{c_1}/J$ (lower) and $T_{c_2}/J$ (upper) as a function 
of $A$. For $A>1$ these are determined using the universal value of the stiffness $\rho/T_c=2/\pi$ 
and the universal value of $\eta=1/4$. The value at the Heisenberg point ($A=1$) was obtained 
from the vorticity results obtained in reference 6. For $A<1$ the data points were taken from 
reference 10.}
\end{figure}

In figure 1 we show the phase diagram for the anisotropic model as a
function of $A$. The data points for $A>1$ were obtained by studying the
finite size behaviour of the structure factor, spin stiffness and
vorticity. Both transitions correspond to vortex unbinding transitions
and we find that for all $A>1$ the correlation length exponent $\eta$
and the ratio of the stiffness to the critical temperature $\rho/T_c$ have the universal
values $\eta=1/4$ and $\rho/T_c=2/\pi$. The data points for $A<1$ have been taken from the 
work of Capriotti et al.\cite{10} who determined the values
of $T_c$ by examining the behaviour of the chirality and the structure factor as a function 
of temperature and lattice size.
The value of $T_c$ at the isotropic point ($A=1$) is that obtained from the vorticity results 
of Southern and Xu.\cite{6}
We have also independently studied the range $-0.5<A<1$ and find some evidence for two 
closely spaced transitions. The upper transition is Ising-like and Monte Carlo snapshots of 
the chirality clearly
indicate Ising-like domains. On the other hand, the spin stiffness exhibits behaviour consistent 
with a vortex unbinding involving the $xy$ spin components but we are unable to directly observe 
vortex excitations in the simulations.
 
The structure of the phase diagram suggests that the Heisenberg point is a multicritical point 
where possibly four phase transition lines meet. For $A>1$ there are two KT transition lines 
whereas as for $A<1$ there is an Ising and a KT line or possibly a single transition line. The 
behaviour of these transition lines as $A\rightarrow 1$ from both
sides strongly suggests that there is a finite temperature phase transition
in the Heisenberg model . In particular, the lower transition for $A>1$ is increasing towards 
the Heisenberg value in this limit. Since the spin
stiffness is zero at large length scales for all finite temperatures when $A=1$, the spin 
correlations decay exponentially. Thus the transition would be quite different from the KT 
transition where spin correlations acquire a power law decay with a 
temperature dependent exponent. The transition at the Heisenberg point
would be purely topological in character. Further theoretical investigations
of this possibility are desirable.

\section{Acknowledgements}                                

This work was supported by the Natural Sciences and Engineering
Research Council of Canada.

\end{document}